\documentclass{emulateapj}

\shorttitle{A new intermediate mass protostar in Cep A HW2}
\shortauthors{J. Mart\'{\i}n--Pintado et al.}

\begin{document}

\title{A new intermediate mass protostar in the Cepheus A HW2 region}

\author{Jes\'us Mart\'{\i}n--Pintado, Izaskun Jim\'{e}nez--Serra, Arturo
Rodr\'{\i}guez--Franco\altaffilmark{1}}
\affil{Dpto. de Astrof\'{\i}sica Molecular e Infrarroja, 
Instituto de Estructura de la Materia,
Consejo Superior de Investigaciones Cient\'{\i}ficas (CSIC),
C/ Serrano 121, E--28006 Madrid, Spain; 
martin@damir.iem.csic.es, izaskun@damir.iem.csic.es,
arturo@damir.iem.csic.es}

\author{Sergio Mart\'{\i}n}
\affil{Instituto de Radioastronom\'{\i}a Milim\'etrica,
Avda. Divina Pastora, Local 20,
E--18012 Granada, Spain; martin@iram.es}

\and

\author{Clemens Thum}

\affil{Institut de Radio Astronomie Millim\'etrique, 300 Rue de la Piscine,
F--38406 St. Martin d'H\`eres, France; thum@iram.fr}

\altaffiltext{1}{Escuela Universitaria de \'Optica,  
Departamento de Matem\'atica Aplicada (Biomatem\'atica),
Universidad Complutense de Madrid,
Avda. Arcos de Jal\'on s/n, E--28037 Madrid, Spain}

\begin{abstract}

We present the discovery of the first molecular hot core associated
with an intermediate mass protostar in the Cep$\,$A$\,$HW2 region. The hot
condensation was detected from single dish and interferometric observations
of several high excitation rotational lines (from 100 to 880$\,$K
above the ground state) of SO$_{2}$ in the ground vibrational state
and of HC$_{3}$N in the vibrationally excited states v$_7$=1 and v$_7$=2. 
The kinetic temperature derived from both molecules is $\sim$160$\,$K. 
The high--angular resolution observations (1.25$''$$\times$0.99$''$)
of the SO$_{2}$ $J=28_{7,21}\rightarrow29_{6,24}$ line (488$\,$K above
the ground state) show that the hot gas is concentrated in a compact 
condensation with a size of $\sim$0.6$''$ ($\sim$430$\,$AU), located
0.4$''$ (300$\,$AU) east from the radio--jet HW2. The total SO$_{2}$ 
column density in the hot condensation is $\sim$10$^{18}$cm$^{-2}$, with 
a H$_{2}$ column density ranging from $\sim$10$^{23}$
to 6$\times$10$^{24}$$\,$cm$^{-2}$. The H$_{2}$ density and the
SO$_{2}$ fractional abundance must be larger than
10$^{7}$$\,$cm$^{-3}$ and 2$\times$10$^{-7}$ respectively. The most
likely alternatives for the nature of the hot and very dense condensation
are discussed. From the large column densities of hot gas, the detection
of the HC$_{3}$N vibrationally excited lines and the large SO$_{2}$ abundance, we
favor the interpretation of a hot core heated by an intermediate mass
protostar of 10$^{3}$$\,$L$_{\odot}$. This indicates that the Cep$\,$A$\,$HW2
region contains a cluster of very young stars.

\end{abstract}


\section{Introduction}

While the general scenario of low mass star formation is reasonably
well understood, the formation of massive stars is still poorly known
(see e.g. \citealt{mck03}). Hot cores represent one of the earliest
phases in massive star formation 
accretion (\citealt{gar99}). They are compact condensations 
($\leq$0.1$\,$pc) with high densities ($\geq$10$^{6}$$\,$cm$^{-3}$), 
high kinetic temperatures ($\geq$100$\,$K) and a very rich chemistry.
The abundances of sulfur--bearing molecules like SO$_{2}$, are expected
to be enhanced in hot cores by several orders of magnitude with respect
to the quiescent gas (\citealt{cha97}). So far, these hot cores
have been mainly detected in objects whose luminosities are larger than 
10$^{4}$$\,$L$_{\odot}$ (\citealt{gar99}). 

Cepheus A (CepA), with a total IR luminosity of
2$\times$10$^{4}$$\,$L$_{\odot}$, is a very active region of massive
star formation. It contains sixteen ultracompact H$\,$II regions, 
molecular outflows and water vapor masers 
(e.g. \citealt{hug84,rod80,torr96,gar96}). The main powering source 
in the region is believed to be the thermal radio--jet, known as HW2, 
which is the brightest of the radio continuum sources  in the
region (\citealt{rod94,gar96}). The surroundings of HW2 show a rather complex 
structure. The morphology of the HW2 radio--jet
suggests the presence of a disk of neutral gas whose size is unknown.  
\citet{torr96} reported evidence for water masers in a rotating and 
collapsing disk around the radio--jet. However, the warm molecular gas (about
50$\,$K) around HW2 shows a ring like structure  with complex kinematics 
which does not seem to be consistent with a simple rotating disk 
(\citealt{torr99}). 

In addition to the expected neutral disk, within a few hundred AU
of the radio--jet, there is clear evidence for very recent 
star formation. Two radio continuum sources detected by \citet{cur02}, 
one of them claimed to be the source of the expanding bubble found
in H$_{2}$O masers (\citealt{torr01}), are very likely protostars.
Furthermore, the  multiple outflow activity (Hayashi, Hasegawa, \&
Kaifu 1988; \citealt{nar96})
seems to be powered by three different young stellar
objects in the surroundings of HW2 (\citealt{cod03}). Two of these sources
have not been yet identified. 

In this letter, we report the detection of a hot (160$\,$K), very
dense ($\geq$10$^{7}$$\,$cm$^{-3}$) compact 
(0.6$''$;$\,$430$\,$AU) condensation located 0.4$''$ (300$\,$AU) 
east from the radio--jet HW2. From the properties of the hot condensation
and the large H$_{2}$ column densities of hot gas, we conclude that
this condensation is likely a hot core associated with an intermediate
mass star with a luminosity of 10$^{3}$$\,$L$_{\odot}$
which could be the driving source of one of the molecular outflows
observed in the CepA region. 

\section{Observations \& Results}

The observations of the high excitation SO$_{2}$ rotational lines
and of the HC$_{3}$N vibrationally excited lines (hereafter HC$_{3}$N$^{*}$)
in Tab.$\,$1 were made with the IRAM%
\footnote{The Institute for Radio Astronomy at Millimeter wavelengths (IRAM)
is supported jointly by the German Max--Planck--Society, the French
Centre National de Recherche Scientifique (CNRS) and the Spanish Instituto
Geogr\'afico Nacional. %
} 30$\,$m telescope at Pico Veleta (Spain). The lines were observed with
the 1.3, 2 and 3$\,$mm SIS receivers simultaneously. The half power beam
width (HPBW) of the telescope ranged from 14$''$
 to 25$''$. The receivers were tuned to single
side band with image rejections larger than 10$\,$dB. The system temperatures
of the 1.3, 2 and 3$\,$mm receivers were, respectively, $\sim$450, $\sim$310
and $\sim$170$\,$K. As spectrometers, we used the VESPA autocorrelator, configured
to provide velocity resolutions ranging from 0.037 to
3.3$\,$km$\,$s$^{-1}$. 
The line intensities were converted
to main brightness temperatures using the beam efficiencies of 0.55,
0.69 and 0.77 for the 1.3, 2 and 3$\,$mm lines respectively.

The $J=28_{7,21}\rightarrow29_{6,24}$ SO$_{2}$ 
line (489 K above the ground state) was also observed with high angular
resolution using the IRAM Plateau de Bure Interferometer --PdBI-- (France)
in the A and B configurations. The 3$\,$mm receiver was tuned to double
side band. The HPBW of the synthesized beam was 1.25$''$$\times$0.99$''$,
PA=--25$^{\circ}$. The correlator was configured to cover 145$\,$MHz 
(439$\,$km$\,$s$^{-1}$) with a spectral resolution of 625$\,$KHz 
(1.89$\,$km$\,$s$^{-1}$). CRL618 (1.73$\,$Jy) and MWC349 (1.03$\,$Jy) 
were used as flux density calibrators. 
The total observing time on--source was about 7$\,$hours. 

Fig.$\,$1 shows the SO$_{2}$ and HC$_{3}$N$^{\ast}$ line profiles
observed with the 30$\,$m telescope and the PdBI toward CepA$\,$HW2, and
Tab.$\,$1 gives the line parameters obtained by fitting Gaussian profiles.
Our $J=16_{2,14}\rightarrow15_{3,13}$  and $J=3_{1,3}\rightarrow2_{0,2}$ SO$_2$ lines 
are in agreement with those of \citet{cod03}. The $J=3_{1,3}\rightarrow2_{0,2}$ SO$_2$ line
will be not used in our analysis since it is easily excited and mainly arises from the 
extended molecular outflow.
Like most of the low excitation lines of many molecules 
(Mart\'{\i}n--Pintado, Bachiller, \& Fuente 1992;
\citealt{ber97,cod03}), all of the high excitation lines of 
SO$_{2}$ and HC$_{3}$N$^{\ast}$ peak around --10$\,$km$\,$s$^{-1}$. 
The SO$_{2}$ lines also show a slightly asymmetric profile with a
weak redshifted wing. This asymmetry is probably associated with the
high velocity shocks observed in SO$_{2}$ (\citealt{cod03}).

Fig.$\,$2 shows the high angular resolution PdBI images of the continuum
emission (thin solid contours and grey map) and of the 
$J=28_{7,21}\rightarrow29_{6,24}$ SO$_{2}$ line intensity at 
--10$\,$km$\,$s$^{-1}$ (thick solid contours). The continuum emission 
is unresolved in our beam with a peak continuum intensity of 
110$\,$$\pm$$\,$15$\,$mJy, in agreement with those previously
measured by Mehringer, Zhou, \& Dickel (1997) and \citet{gom99}, and consistent
with the continuum emission arising from the HW2 radio--jet. For all
radial velocities, the $J=28_{7,21}\rightarrow29_{6,24}$ SO$_{2}$
line emission is also spatially unresolved, indicating that
the highly excited gas is very compact with a size of $\leq$0.6$''$
(430$\,$AU). The unresolved SO$_{2}$ condensation contains most of the flux
(70\%) measured with the 30$\,$m telescope. As shown in Fig.$\,$2, the 
SO$_{2}$ emission is located 0.4$''$ east from the
radio--jet. This is consistent with the warm 
NH$_{3}$ emission observed in the (3,3) line by \citet{torr99}.

\section{Properties of the compact molecular condensation}

Tab.$\,$2 summarizes the properties of the compact molecular condensation.
Fig.$\,$3 shows the population diagrams 
(see e.g. \citealt{gol99}) derived for SO$_{2}$ and HC$_{3}$N$^{\ast}$. The level 
populations of both molecules can be fitted by straight
lines with similar excitation temperatures of 150--160$\,$K, indicating the presence 
of a hot condensation (hereafter
HC). 
We have constrained the size of the HC by using the upper 
limit to the size measured from the interferometric
maps of 0.6$''$, and the lower limit of 0.3$''$ derived from the
measured main beam brightness temperature of the 
$J=28_{7,21}\rightarrow29_{6,24}$ SO$_{2}$ line and the excitation
temperature of 160$\,$K. In the following
discussion, we will consider a size of 0.6$''$ (430$\,$AU) for the HC. 
 The derived 
excitation temperature from the HC$_{3}$N$^{\ast}$ transitions is unlikely
affected by optical depth effects since the observed lines are weaker than those 
expected from a source  of 0.6$''$ and optically thick emission with an excitation temperature of 160 K. 

 Combining the column density derived from the PdBI data of the $J=28_{7,21}\rightarrow29_{6,24}$ SO$_{2}$ line 
(see Table 1) with the estimated source size of 0.6$''$  and the excitation temperature of 160 K, we derive
a total SO$_{2}$ column density for the HC of $\sim$10$^{18}$$\,$cm$^{-2}$.
Although the H$_2$  column density of the HC
is unknown, it can be constrained 
between $10^{23}$ and $6\times10^{24}\,$cm$^{-2}$. The upper limit
of $6\times10^{24}\,$cm$^{-2}$ is derived from our 3$\,$mm continuum
intensity of 30$\,$mJy measured at the location of the HC, and by
assuming that it arises from dust at a temperature of 160$\,$K. The lower limit 
is obtained by assuming the extreme case that 50\% of the cosmic sulfur 
abundance is in SO$_{2}$. Assuming spherical symmetry, 
the H$_{2}$ densities in the HC range between
10$^{7}$ and 6$\times$10$^{8}$$\,$cm$^{-3}$. Hereafter 
(see Tab.$\,$2), we will consider the intermediate case 
for a H$_{2}$ column density of 10$^{24}$$\,$cm$^{-2}$ (i.e. a SO$_{2}$ abundance
of 10$^{-6}$). In this case, the H$_{2}$ density and the mass of the
HC will be $\sim$10$^{8}$$\,$cm$^{-3}$ and 4$\times$10$^{-2}$M$\odot$
respectively.  The kinetic temperature, (T$_{\rm kin}$), of the HC is unknown, but it must
be equal or larger than the derived excitation temperature of 160 K. For
the high density ($>$10$^7$$\,$cm$^{-3}$) derived for the HC, the SO$_{2}$ transitions are thermalized and the
excitation temperature should be a direct measurement of the kinetic temperature. Furthermore
the dust and gas must be also closely coupled and the dust temperature should be
equal to the gas kinetic temperature. Since the HC is optically
thick in the mid--IR, we estimate a bolometric luminosity of 
$\sim$10$^{3}$$\,$L$_{\odot}$ by assuming that it emits like a black
body at 160$\,$K. 

In summary, we have detected a warm (160$\,$K), very dense 
($>$10$^{7}$$\,$cm$^{-3}$) and compact condensation (size
$\sim$400$\,$AU) located 300$\,$AU east from the radio--jet HW2. 
The HC has a total mass of 4$\times$10$^{-2}$$\,$M$_{\odot}$, a luminosity of 
$\sim$10$^{3}$$\,$L$_{\odot}$ and a very large abundance 
($\geq$2$\times$10$^{-7}$) of SO$_{2}$.

\section{The nature of the hot condensation. A hot core}

In view of the complexity of the region in the vicinity of HW2, we
discuss the two most likely possibilities for the nature of the HC. The
first one consists of a hot and dense spot in the expected circumstellar disk
around the radio--jet. In the disk scenario, the large H$_{2}$ column density 
(10$^{24}$$\,$cm$^{-2}$) of hot gas (160$\,$K) in the HC 
could be externally heated by the IR radiation from HW2 or by the shocks 
produced by the stellar wind of HW2. For the
projected distance between HW2 and the HC, external heating by radiation
would require the IR luminosity of HW2 to be few 10$^5$$\,$L$\odot$
(Fig.$\,$7 of Kaufman, Hollenbach, \& Tielens 1998), which is at least one
order of magnitude larger than that measured in the whole region 
of 2$\times$10$^{4}$$\,$L$\odot$. Shock heating would require 
collisional excitation of
the v$_{7}$=1 and 2 vibrationally levels of HC$_3$N and the conversion of 10\% 
of the total bolometric luminosity in the region into mechanical
luminosity. Collisional excitation of the vibrational levels of HC$_3$N
is possible since the H$_{2}$ densities estimated for the HC are 
close to the critical densities of 2$\times$10$^{8}$$\,$cm$^{-3}$ 
(\citealt{gol82}). Shocks should then provide a mechanical luminosity 
of 10$^{3}$L$\odot$ in a region of only 400$\,$AU. This is very
unlikely since the total mechanical luminosity in the large scale
outflow powered by HW2 is only 30$\,$L$\odot$ (\citealt{nar96}), 
nearly two orders of magnitude smaller than the mechanical luminosity 
required to heat the HC.

The second possibility is that we are observing a hot core internally
heated. This option will explain the presence of the vibrationally
excited emission which in hot cores is usually produced by IR radiation
(see e.g. \citealt{vic00}). The luminosity and the mass of
the hot core in HW2 will be more than one order of magnitude smaller
than those reported for hot cores associated with massive stars. Then, 
the HC will be the first hot core associated with an intermediate
mass star. The luminosity of the hot core would require a main sequence
B5 star. Large SO$_{2}$ abundance ($\sim$10$^{-7}$) seems to be
a general property of hot cores associated with massive protostars
(\citealt{kea01}). Our lower limit to the SO$_{2}$ abundance of
2$\times$10$^{-7}$ suggests that intermediate mass hot cores have
similar sulfur chemistry than those associated with massive protostars.
The derived HC$_{3}$N abundance in the HC of $\sim$10$^{-8}$, is also
in agreement with the abundances measured in hot cores associated
with massive stars (\citealt{vic00}). It has been proposed that the large
abundance of sulfur--bearing molecules can be driven by fast gas phase
reactions after the evaporation from grain mantles of key molecules 
like H$_{2}$S (\citealt{cha97}), OCS (\citealt{vdt03,smar05})
and SO (\citealt{jim05}) or atomic sulfur (\citealt{wak04}). 
From the available data on other
sulfur--bearing molecules, SO$_2$ seems to be the most abundant  
species in the HC. We can set upper limits to the abundance of other sulfur--bearing
molecules and of atomic sulfur by assuming that all their emissions observed
with beams of 10--30$''$, are optically thin and arise from the hot
core. From the H$_{2}$S, SO and [S$\,$I]$\,$25.2$\mu$m
observations of \citet{cod03}, \citet{mar92} and \citet{wri96}
respectively, we derive column densities for the
three species of 1--2$\times$10$^{17}$cm$^{-2}$. Their corresponding
abundances are $\leq$10$^{-8}$, at least one order of magnitude smaller
than the SO$_{2}$ abundance. \citet{viti04} have performed model
calculations to follow the sulfur chemistry evolution from the collapse phase
to the hot core stage using new experimental data on desorption temperatures
for mixed ices. The model predicts that 
 an intermediate mass star
of 5$\,$M$\odot$ in the hot core phase will evaporate H$_{2}$S from the
grain mantles, and after few 10$^{5}$ years, will produce SO$_{2}$ in
gas phase with abundances similar to that observed in the intermediate 
mass hot core in HW2. 

The discovery reported in this letter of an intermediate mass protostar,
and the recent detection of two embedded young stellar objects (\citealt{cur02})
within 500$\,$AU of HW2, support the scenario of the formation of a cluster
of stars. The new protostar in the HW2 region is an excellent candidate
for powering one of the three molecular outflows observed in the region
(\citealt{cod03}). If the relationship found between the mechanical
luminosity of the molecular outflows and the luminosity
of their exciting sources (\citealt{she96}) holds for the outflows
in the CepA region, the new protostar will be luminous enough to power
the east outflow (\citealt{hay88}). However, high angular resolution
observations of these outflows are needed to confirm if the new protostar
is the driving source of one of the multiple outflows found in this
region.

In summary, we have detected a hot and dense condensation in the vicinity
of the radio--jet HW2 in the CepA region. From the properties of the
hot condensation, we suggest that it is a hot core heated by an intermediate
mass protostar. This protostar could be the driving source of one
of the observed multiple molecular outflows. The detection
of a new protostar in the vicinity of HW2 supports the idea of the
formation of a cluster of massive and intermediate stars in the
CepA region.

\acknowledgments

This work has been supported by the Spanish MEC under projects
number AYA2002--10113--E, AYA2003--02785--E and ESP2004--00665.

\begin{figure}
\epsscale{1.0}
\plotone{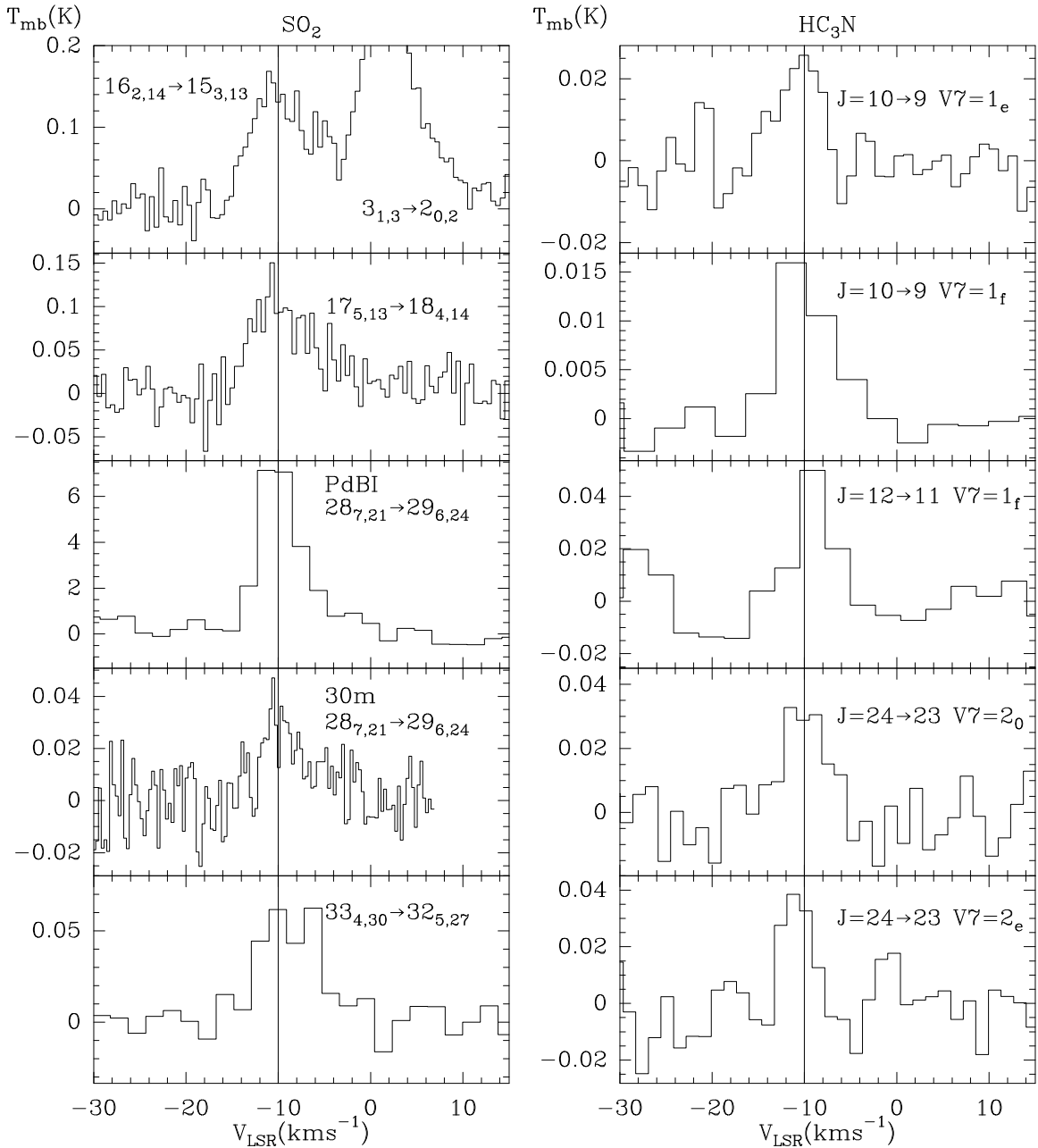}
\caption{Sample of profiles of the SO$_{2}$ and HC$_{3}$N vibrationally
excited rotational transitions observed toward CepA$\,$HW2.}
\label{fig1}
\end{figure}

\begin{figure}
\epsscale{1.0}
\plotone{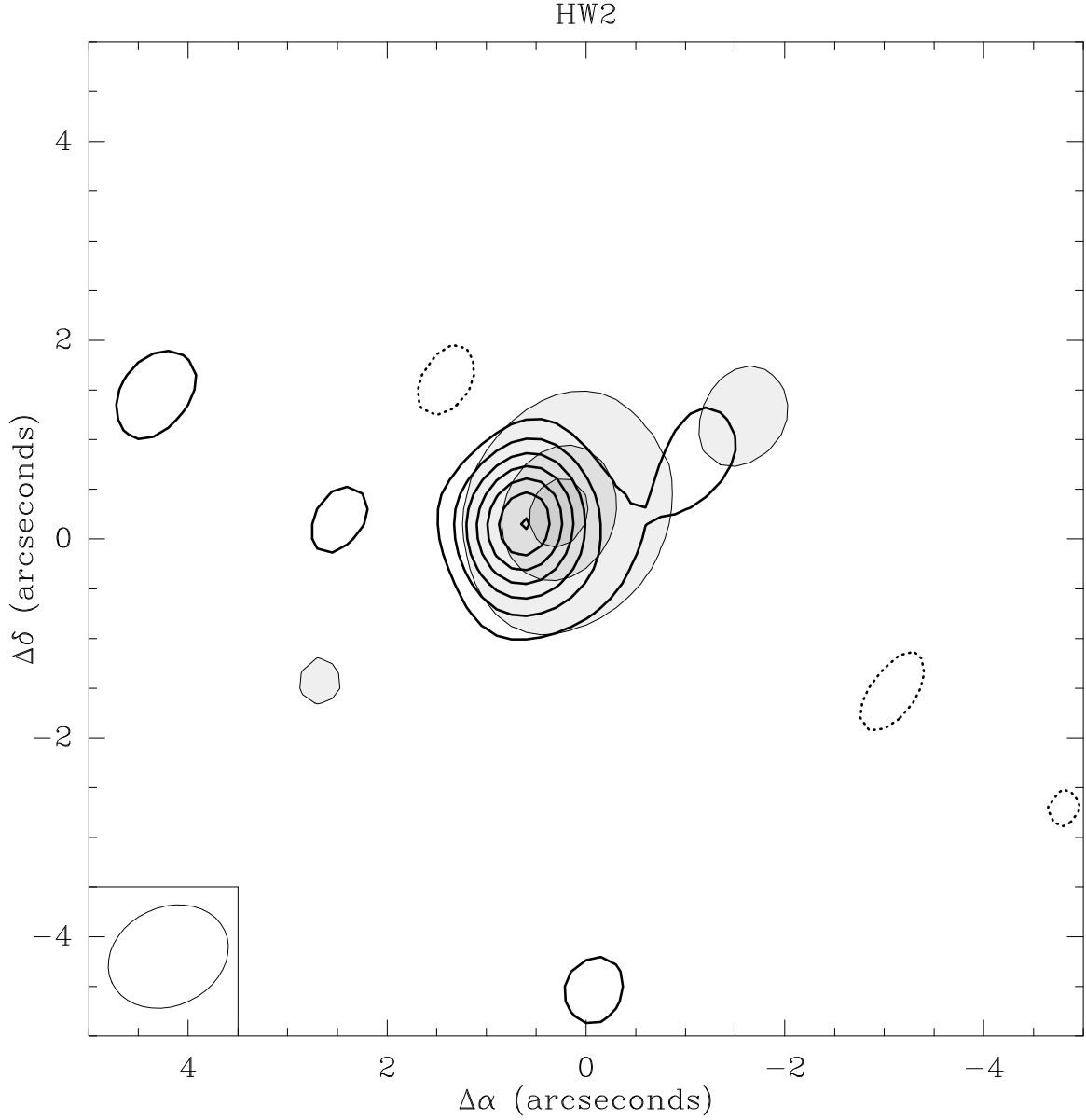}
\caption{Images of the $J=28_{7,21}\rightarrow29_{6,24}$ SO$_{2}$
line intensity at --10$\,$km$\,$s$^{-1}$ (thick solid contours) 
superimposed on the radio continuum emission at 3$\,$mm (thin solid 
contours and grey map). The lowest level and the step for the continuum
map are  0.01$\,$Jy/beam and 0.04$\,$Jy/beam respectively. The contour
levels for the SO$_{2}$ map are --0.01, 0.01, 0.02, 0.03, 0.04,
0.05, 0.06 and 0.07$\,$Jy/beam$\,$km$\,$s$^{-1}$. Beam size is shown
in the lower left corner.}
\label{fig2}
\end{figure}

\begin{figure}
\epsscale{1.0}
\plotone{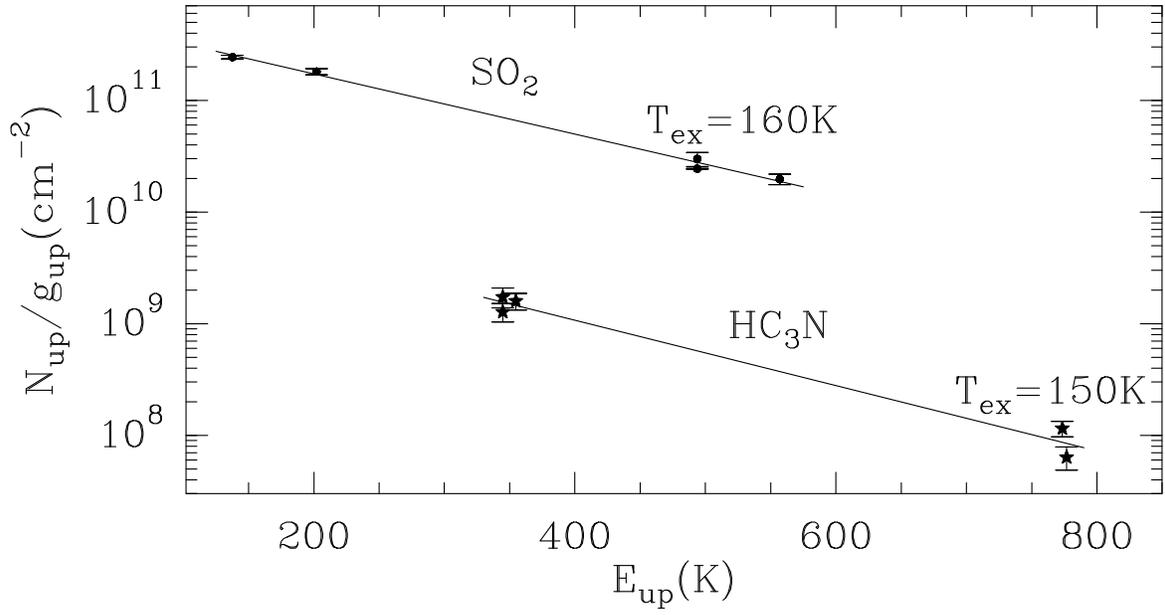}
\caption{Population diagrams for all the SO$_{2}$ and HC$_{3}$N$^{*}$
rotational transitions observed toward CepA$\,$HW2. The column densities are 
averaged values in a beam of 24$''$ for SO$_2$ and 27$''$ for HC$_{3}$N$^{\ast}$.}
\label{fig3}
\end{figure}

\clearpage

\begin{deluxetable}{lrccc}
\tabletypesize{\scriptsize}
\tablecaption{Observed parameters and derived column densities 
for SO$_2$ and HC$_3$N in CepA$\,$HW2.
\label{tbl-1}} 
\tablewidth{0pt}
\startdata

\\ \hline\hline 

Line & V$_{LSR}$\tablenotemark{a} & $\Delta v$\tablenotemark{a}& T$_{mb}$\tablenotemark{a} & $N_{up}$ \\
& (km$\,$s$^{-1}$) & (km$\,$s$^{-1}$) & (K) &
($\times$10$^{11}$$\,$cm$^{-2}$) \\

\hline

\multicolumn{5}{c}{{\bf SO$_2$}}\\
$16_{2,14}\rightarrow15_{3,13}$ & -9.7 (1) & 7.9 (4) & 0.144
(3) & 80.4 \\
$17_{5,13}\rightarrow18_{4,14}$ & -9.6 (3) & 7.8 (6) & 0.100
(3) & 70.3 \\
$28_{7,21}\rightarrow29_{6,24}$ & -10.0 (1) &
5.2 (3) & 8.8 (1) & 3628\tablenotemark{b} \\
$28_{7,21}\rightarrow29_{6,24}$ & -9.4 (5) & 6 (1) &
0.030 (4) & 15.4 \\
$33_{4,30}\rightarrow32_{5,27}$ & -8.8 (4) & 7.8 (9) & 0.060 (4)
& 30.0 \\
\multicolumn{5}{c}{{\bf HC$_3$N}} \\
$10\rightarrow9~v_7=1e$ & -10.5 (5) & 4.8 (9) & 0.026 (6) &
0.36 \\
$10\rightarrow9~v_7=1f$ & -10.3 (5) & 5 (2) & 0.017 (4) &
0.27 \\
$12\rightarrow11~v_7=1f$ & -8.9 (5) & 4.3 (8) & 0.050
(3) & 0.57 \\
$24\rightarrow23~v_7=2_0$ & -10.2 (5) & 7 (1) & 0.04 (1)
&  0.33 \\
$24\rightarrow23~v_7=2_e$ & -10.8 (4) & 3.4 (8) & 0.04 (1)
&  0.18 \\ 

\enddata
\tablenotetext{a}{The 1$\sigma$ errors from the gaussfits are shown in parenthesis.}
\tablenotetext{b}{SO$_2$ column density derived for a beam of 
1.25$''$$\times$0.99$''$.}
\end{deluxetable}

\begin{deluxetable}{cccccccccc}
\tabletypesize{\scriptsize}
\tablecaption{Physical properties of the compact molecular condensation HC.
\label{tbl-2}} 
\tablewidth{0pt}
\startdata

\\ \hline\hline
Source & Location\tablenotemark{a} & Size & T$_{kin}$ &  n$_{{\rm H}_2}$ &
\multicolumn{3}{c}{Column densities\tablenotemark{b}} &  Luminosity &
Mass \\
& ($''$) & ($''$) & (K) & (cm$^{-3}$) & SO$_2$ & HC$_3$N &
H$_2$\tablenotemark{c} & (L$_\odot$) & (M$_\odot$) \\

\hline

 HC & $0.32(0.03)\times0.15(0.03)$ & 0.3--0.6 & 160(10) & 1.5$\times10^{8}$ &
 1$\times$10$^{18}$ & 3$\times10^{16}$ & 1$\times$10$^{24}$ & 1$\times$10$^{3}$
 & 4$\times10^{-2}$ 

\tablenotetext{a}{Offset from the continuum peak at $\alpha(2000)=22^{\rm h}
56^{\rm m} 17.97^{\rm s}(0.01)$, $\delta(2000)=62^{\rm o}
1' 49.70''(0.08)$ derived from gaussian fits.}
\tablenotetext{b}{For a source size of $0.6''$ ($430\,$UA) and a distance to the
source of 725$\,$pc. Column densities are in units of cm$^{-2}$.}
\tablenotetext{c}{See text for details.}

\enddata

\end{deluxetable}

\end{document}